\documentclass{iopart}
\usepackage{tabularx, graphicx}

\newcommand{\adl}{a^{\dagger}_{l,k}}
\newcommand{\adr}{a^{\dagger}_{r,-k}}
\newcommand{\al}{a_{l,k}}
\newcommand{\ar}{a_{r,-k}}

\begin{document}

\title{Sudden interaction quench in the quantum sine-Gordon model}

\author{Javier Sabio$^{1,2}$ and Stefan Kehrein$^3$}

\address{$^1$Instituto de Ciencia de Materiales de Madrid
(CSIC), Sor Juana In\'es de la Cruz 3, E-28049 Madrid, Spain Institute of Physics\\$^2$Departamento de F{\'\i}sica de Materiales. Universidad
  Complutense de Madrid. 28040 Madrid. Spain.\\$^3$ Arnold-Sommerfeld-Center for Theoretical Physics, Center for NanoSciences and Department f\"ur Physik, Ludwig-Maximilians-Universit\"at M\"unchen, Theresienstrasse 37, 80333 M\"unchen, Germany}
\ead{javier.sabio@icmm.csic.es}

\begin{abstract}
We study a sudden interaction quench in the weak-coupling regime of the quantum sine-Gordon model.
The real time dynamics of the bosonic mode occupation numbers is calculated using the flow equation method.
While we cannot prove results for the asymptotic long time limit, we can establish the existence
of an extended regime in time where the mode occupation numbers relax to twice their equilibrium values.
This factor two indicates a non-equilibrium distribution and
is a universal feature of weak interaction quenches. The weak-coupling quantum sine-Gordon model
therefore turns out to be on the borderline between thermalization and non-thermalization. 
\end{abstract}

\maketitle

\section{Introduction}

The last years have witnessed an increasing interest in the dynamics of isolated quantum many-body systems. As it has happened to many other fields that were considered mainly academical in the past, the recent advances in experiments with ultracold gases confined in optical lattices have opened up the possibility of making real tests of the long time evolution of essentially isolated quantum systems. And following it, a pletora of unanswered questions have regained the attention of the scientific community. 

One important issue in this context is the question of thermalization in quantum many-body systems.
Coupled to environments, the dynamics of quantum systems is known to lead to equilibrium states described by the thermal ensembles of Quantum Statistical Mechanics. But once the system is isolated and initialized in a highly non-thermal state, it remains unclear if the final, long-time state of the system can be described with one of those states, i.e., if the system has thermalized. Rigorously, it is easy to show that Quantum Mechanics does not allow evolution from a pure state to a thermal distribution, as unitary time evolution preserves the purity. However, to which extent thermal averages reproduce long-time quantum averages, and the conditions whereby this happens, can still be considered an open problem \cite{Deutsch91, Srednicki94, Rigol08, Cramer08, Reimann08, Moeckel08, Moeckel09, Hackl09, Kehrein09, Eckstein08, Eckstein09,Manmana07,Kollath07,Schollwoeck08}.  

From this debate, new questions arose concerning the role that integrability plays in the long-time evolution of quantum systems. The dynamics of integrable systems is expected to be very constrained due to the large number of constants of motion, and hence ordinary thermalization should not be present in such systems. Experimental studies of systems near integrability point in this direction \cite{Kinoshita06}. However, it has been argued that integrable systems could still relax to an ensemble described by quantum statistical mechanics if all the constants of motion are included as constraints \cite{Rigol07}. In this respect the thermalization debate is still relevant even regarding integrable systems \cite{Calabrese07, Cazalilla06}.

Similar to non-equilibrium classical statistical mechanics, the lack of a general framework to study non-equilibrium quantum problems makes it necessary to focus on specific models with the aim of extracting general features from them. However, long-time evolution of far-from-equilibrium systems is a very challenging topic itself, and only recently suitable techniques, both analytical and numerical have been developed. In this paper we employ the 
{\it forward-backward} scheme \cite{Hackl08,HacklKehrein09} based on the flow equation method \cite{Wegner94,KehreinBook}, which has turned out to  
be a realiable and powerful approach to solve Heisenberg equations of motion for operators. The main idea is to use 
the flow equation method to diagonalize the Hamiltonian in a controlled approximation, and then to study 
the time evolution problem in this diagonal basis. Hence real time evolution in this basis becomes simple
and can be extended to long times without secular terms. All the difficulties of the problem are therefore 
encoded in the unitary transformation.

In this paper we apply this approach to study the real time dynamics of the sine-Gordon model after a sudden quench of the interaction. Sudden interaction quenches are very interesting for studying non-equilibrium dynamics since they provide far-from-equilibrium initial states, but simplify the theoretical calculations since this initial state is simple. In optical lattices it is also possible to implement them experimentally \cite{Greiner}. 

The quantum sine-Gordon model is a $1+1$ dimensional scalar field theory with a rich phase diagram. It is an integrable model, whose exact solution can be obtained by using the Bethe Ansatz \cite{Zamolodchikov77}. However, as usual this
exact solution does not guarantee a simple calculation of the observables.  Therefore it is interesting and necessary 
to implement approximate schemes like the forward-backward method. The choice of the sine-Gordon model
is motivated by the fact that it is a paradigm for one dimensional translation-invariant interacting systems, with
mappings that connect it to many other models. For example, it arises as the effective description of a system of interacting fermions in one dimension with backscattering, or for spin-1/2 quantum spin chains \cite{Giamarchi03}. 
Sudden quenches have been already studied for one-dimensional fermions with {\it density-density interactions}, i.e., in the context of the Luttinger model \cite{Cazalilla06}: this is an integrable model whose exact solution is quadratic once it is written in terms of bosonic excitations \cite{Giamarchi03}. Likewise, the interaction 
quench to the Luther-Emery line of the sine-Gordon model leads to a quadratic model expressed in 
fermions \cite{Cazalilla08}. However,
in general such simple mappings are not possible for the sine-Gordon model: despite being integrable, it cannot be expressed as a quadratic Hamiltonian, and hence one can expect a redistribution of energy between different modes.  

The paper is organized as follows. In section~2 we introduce the main features of the model including its phase diagram and we describe its solution by the flow equation method. In section~3 we implement the forward-backward scheme to study the time evolution of the mode occupation operator in the weak-coupling region of the phase diagram. In section~4 the resulting expressions are employed to analyze the effect of a sudden quench of interactions: these are the central results of this work. 

\section{The model and the flow equation solution}

The sine-Gordon model is an ubiquitous model widely studied in many different areas of physics. Its classical
$1+1$~dimensional version became very popular in the 1970s as it has  
non-perturbative solutions known as solitons \cite{GogolinBook}. 
Here we will be interested in its quantized counterpart, namely the quantum sine-Gordon model. As mentioned before
the quantum sine-Gordon model is related to one-dimensional fermions with backscattering. 
There are many other similar mappings like to the one-dimensional Hubbard model near half-filling, the Coulomb gas problem, 
the two-dimensional classical $X-Y$ model, and quantum spin chains \cite{Giamarchi03}. Therefore the quantum
sine-Gordon model is a natural and important setting for understanding quench dynamics.

The Hamiltonian of the model is defined as follows:
\begin{equation}
{\mathcal H} =  \int dx \left( \frac{1}{2} \Pi^2(x) + \frac{1}{2} (\frac{\partial \phi}{\partial x})^2 + \frac{g}{2 \pi a^2}\cos(\beta \phi(x))\right)
\end{equation}
where $\phi(x)$ is a scalar bosonic field and $\Pi(x)$ its conjugate momentum field. In order to impose the quantum structure, they must satisfy the commutation relations:
\begin{equation}
[\Pi(x), \phi(y)] = -i \delta(x-y)
\end{equation} 
The Hamiltonian contains the parameter $\beta$ and the coupling constant $g$, which define the phase diagram of the model. In the sequel we will usually use the parameter $\alpha^2 = \beta^2/4\pi$, which will turn out to be directly related to the scaling dimension of the $\cos$-interaction term. The rest of the parameters are used to regularize the theory: $a$ is a lattice discretization parameter (its inverse $1/a$ plays the role of an ultraviolet cutoff) and $L$ is the system size (its inverse is the infrared cutoff). 

\begin{figure}
\begin{center}
\includegraphics[width=3.5in]{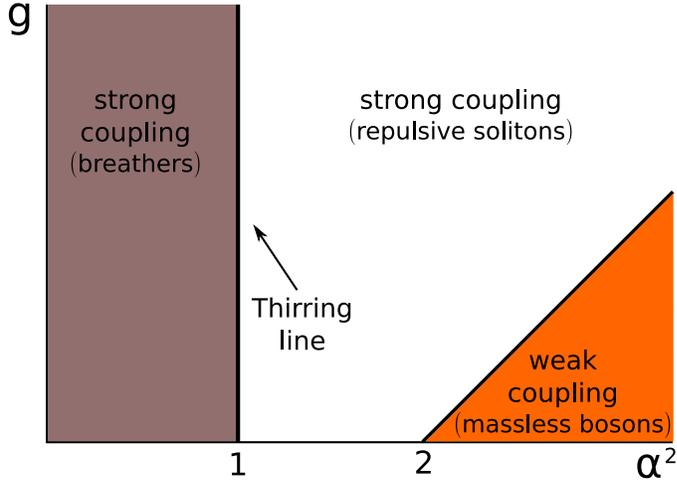}
\caption{Schematic phase diagram of the quantum sine-Gordon model}
\label{fig1}
\end{center}
\end{figure}

A schematic picture of the phase diagram of the model is shown in Fig. \ref{fig1}. At $\alpha^2 = 1$ (Thirring line) the Hamiltonian can be mapped to the non-interacting Thirring model \cite{Coleman}, 
whose relevant degrees of freedom are fermions \cite{Mandelstam75}, which can be identified with quantized solitons of the 
sine-Gordon equation. Away from the Thirring line these fermions experience an interaction:
the region $\alpha^2 > 1$ corresponds to repulsive quantum solitons, whereas in the region $\alpha^2 < 1$ 
the interaction is attractive, which leads to bound states called breathers. 

\begin{figure}
\begin{center}
\includegraphics[width=3.5in]{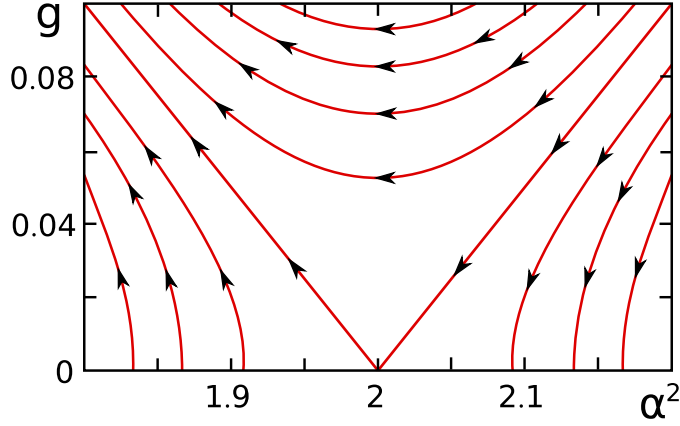}
\caption{Kosterlitz-Thouless like renormalization flow close to $\alpha^2 = 2$}
\label{fig2}
\end{center}
\end{figure}

Another interesting point of the phase diagram occurs near $\alpha^2 = 2$. Here the system undergoes a Kosterlitz-Thouless continuous phase transition, which can be understood from the renormalization group equations for the flowing coupling constants \cite{Kosterlitz72}:
\begin{eqnarray}
\frac{d g}{d \log \Lambda} = (\alpha^2-2) g\\
\frac{d \alpha^2}{d \log \Lambda} = \alpha^4 g^2
\end{eqnarray}
where $\Lambda = 1/\sqrt{2\pi}a$ is the ultraviolet cutoff. A graphical solution of these equations is shown in Fig. \ref{fig2}. For $\alpha^2 < 2$ the coupling constant $g$ flows to strong coupling, which signals the opening of a gap in the spectrum. This corresponds to the emergence of massive fermionic solitons as the appropriate low-energy degrees of freedom for the model. For $\alpha^2 > 2$ we have the weak-coupling regime, where the coupling constant $g$ flows to zero, and the relevant degrees of freedom are massless bosons. In this region, an approximate solution of the equations 
for fixed initial parameter $\alpha_0$ in the weak-coupling limit $|g_0|\ll 1$ is simply:
\begin{eqnarray}
g(\Lambda) \simeq g_0 (\frac{\Lambda}{\Lambda_0})^{\frac{2-\alpha_0^2}{2}} \label{weakflow1}\\
\alpha^2(\Lambda) \simeq \alpha_0^2 +O(g_0^2)
\label{weakflow2}
\end{eqnarray}

In this paper we will be mainly interested in this so-defined weak-coupling limit
where the flow of $\alpha^2$ can be neglected to leading order.  

Next we express the fields in terms of their bosonic modes:
\begin{eqnarray}
\phi(x) = - \frac{i}{\sqrt{4 \pi}} \sum_{k>0} \frac{e^{-\frac{k a}{2}}}{\sqrt{k}}\left(e^{-ikx}(a_{l,k}^{\dagger}+ a_{r,-k}) - e^{i k x}(a_{l,k} + a_{r,-k}^{\dagger}) \right)\\
\Pi(x) = \frac{1}{\sqrt{4 \pi}} \sum_{k>0} e^{-\frac{k a}{2}} \sqrt{k} \left( e^{-ikx}(a_{l,k}^{\dagger}-a_{r,-k}) + e^{i k x}(a_{l,k} - a_{r,-k}^{\dagger}) \right)
\end{eqnarray}
The various creation and annihilation operators obey the usual bosonic commutation relations:
\begin{equation}
[a_{l,k},a_{l,k'}^\dagger]=[a_{r,-k},a_{r,-k'}^\dagger]=\delta_{kk'}\,\frac{L}{2\pi} \quad \forall k,k'>0
\end{equation}
We use the Mandelstam vertex operators \cite{Mandelstam75} to construct the Hamiltonian:
\begin{eqnarray}
V_l(\alpha;x) \equiv :\exp(\alpha \sum_{k>0} e^{-ak/2}\frac{1}{\sqrt{k}}(e^{-ikx}\adl - e^{ikx}\al)): \nonumber\\ =(\frac{L}{2\pi a})^{\alpha^2/2} \exp(\alpha \sum_{k>0} e^{-ak/2}\frac{1}{\sqrt{k}}(e^{-ikx}\adl - e^{ikx}\al))\\
V_r(\alpha;x) \equiv :\exp(\alpha \sum_{k>0} e^{-ak/2}\frac{1}{\sqrt{k}}(e^{-ikx}\ar - e^{ikx}\adr)):  \nonumber\\ =(\frac{L}{2\pi a})^{\alpha^2/2} \exp(\alpha \sum_{k>0} e^{-ak/2}\frac{1}{\sqrt{k}}(e^{-ikx}\ar - e^{ikx}\adr))
\end{eqnarray}
where $:{\mathcal O}: = {\mathcal O} - \langle 0|{\mathcal O}| 0 \rangle$ means normal ordering of the operator ${\mathcal O}$ with respect to the non-interacting ground state \cite{vonDelft98}. The Hamiltonian now reads:
\begin{eqnarray}
\mathcal{H} = \sum_{k>0} k (\adl \al + \adr \ar) \nonumber \\+ \frac{g}{2\pi a^2}(\frac{2\pi a}{L})^{\alpha^2}\int dx (V_l(\alpha;x) V_r(-\alpha;x) + V_r(\alpha;x) V_l(-\alpha;x))
\end{eqnarray}
This Hamiltonian can be studied in its entire phase diagram using the flow equation approach \cite{KehreinBook} as shown in Refs. \cite{Kehrein99,Kehrein01}. The key idea behind this method is to successively apply infinitesimal unitary transformations that eventually diagonalize the Hamiltonian. The  stable choice of such a transformation sequence requires energy scale separation similar to conventional renormalization approaches: as the transformation progresses, more and more
energy-diagonal interaction matrix elements are eliminated. Such a scheme was proposed by Wegner \cite{Wegner94} and independently by Glazek and Wilson \cite{Glazek93, Glazek94}. Wegner showed that a suitable infinitesimal transformation is obtained with the following {\it canonical} generator:
\begin{equation}
\eta(B) = [{\mathcal H}_0(B), {\mathcal H}_{int}(B)]
\end{equation} 
where ${\mathcal H}_0$ is the diagonal and ${\mathcal H}_{int}$ is the interaction part of the Hamiltonian.
$B$ is the {\it flow parameter} that parametrizes the diagonalizing flow. It can be related to an energy scale $\Lambda_B = 1/\sqrt{B}$ in analogy to the conventional RG scheme. The key difference is that the flow equation
$\Lambda_B$ corresponds to an energy difference that is being eliminated, whereas in a conventional RG scheme $\Lambda$
correponds to the UV-cutoff, which is an absolute energy scale.

The flow of the Hamiltonian is given by the following differential equation:
\begin{equation}
\frac{d {\mathcal H}(B)}{d B} = [\eta(B), {\mathcal H}(B)]
\end{equation} 
The methodological challenges come from the implementation of such a transformation and the integration of the ensuing differential equations. In most cases approximations are required in order to get a closed sets of equations. However, these approximations do not necessarily match those employed in other methods like perturbation theory, which, e.g., opens the possibility to access non-perturbative regimes using flow equations. For a comprehensive review of the flow equation method and its applications we refer the reader to Ref.~\cite{KehreinBook}.

Before studying the non-equilibrium dynamics of the sine-Gordon model, we first briefly review 
the flow equation solution of the equilibrium sine-Gordon model in order to make this
paper self-contained. More details of this calculation can be found in Refs.~\cite{Kehrein99,Kehrein01}.
 
It turns out to be more convenient to work with Fourier transformed vertex operators:
\begin{eqnarray}
V_l(-\alpha; k) \equiv \frac{1}{2\pi} \int dx e^{-i k x} V_l(-\alpha;x) \nonumber \\
V_l(\alpha;k) \equiv V_l^\dagger(-\alpha;k) = \frac{1}{2\pi} \int dx e^{i k x} V_l(\alpha;x) \\
V_r(-\alpha; k) \equiv \frac{1}{2\pi} \int dx e^{-i k x} V_r(-\alpha;x) \nonumber\\
V_r(\alpha;k) \equiv V_r^\dagger(-\alpha;k) = \frac{1}{2\pi} \int dx e^{i k x} V_r(\alpha;x) 
\end{eqnarray}
Some relevant properties of these operators are summarized in Appendix A. In this representation the generator of the unitary transformation consists of two parts:
\begin{eqnarray}
\eta(B) = \eta^{(1)}(B) +  \eta^{(2)}(B) \nonumber\\
\eta^{(1)}(B) = 8 \pi^2 \sum_p p\, u(p;B) \left( V_l(\alpha; p) V_r(-\alpha;p) - h.c.\right) \\
\eta^{(2)}(B) = -\psi(B) \sum_{k>0}(\adl \adr - \ar \al)
\end{eqnarray}
where:
\begin{eqnarray}
u(p;B) = \frac{g(B)}{(2\pi a)^2} (\frac{2 \pi a}{L})^{\alpha^2} e^{-4 p^2 B} \\
\psi(B) = -\frac{32}{a^2}(\frac{32B}{a^2})^{1-\alpha^2(B)}g^2(B)\frac{\alpha^2(B)}{4\Gamma(\alpha^2(B)-1)}
\end{eqnarray}
The actual values for these coefficients are obtained by solving the flow equations
for $g(B)$ and $\alpha(B)$ \cite{Kehrein99,Kehrein01}: 
\begin{eqnarray}
\frac{d \alpha^2}{d l} = \frac{\alpha^4(g^2 + {\mathcal O}(g^3))}{4\pi\Gamma(\alpha^2 -1)}\\
\frac{d g}{d l} = (\alpha^2 -2) g + {\mathcal O}(g^2) 
\end{eqnarray}
with $l \equiv -\frac{1}{2}\log(32B/a^2)$. This flow succesfully describes the different regions of the quantum sine-Gordon phase diagram, from the weak-coupling regime close to the Kosterlitz-Thouless transition, where excitations are massless bosons, to the Thirring line at $\alpha^2= 1$ with massive solitonic excitations.

Since we are interested in the weak-coupling limit ($g_0$ small for fixed initial $\alpha$)
in this paper, we can neglect the flow of $\alpha(B)$ to leading order
and identify $\alpha(B)$ with its initial value~$\alpha$.
The effective diagonal Hamiltonian generated in the limit $B=\infty$ then has the following form:
\begin{eqnarray}
H(B = \infty) = H_0 + H_{diag}(B=\infty) \label{diagonal}\\
H_0 = \sum_{k>0} k (a_{l,k}^{\dagger} a_{l,k} + a_{r,-k}^{\dagger} a_{r,-k})\\
H_{diag}(B = \infty) = \sum_{k>0} \omega_k(B = \infty) 
\label{defHdiag} \\
\qquad\times\big(P_l(\alpha;-k)P_l^{\dagger}(\alpha;-k) + P_l^{\dagger}(\alpha;k) P_l(\alpha;k) \nonumber \\
\qquad + P_r^{\dagger}(\alpha;-k) P_r(\alpha;-k) + P_r(\alpha;k) P_r^{\dagger}(\alpha;k)\big) \nonumber
\end{eqnarray}
Here
\begin{eqnarray}
P_j(\alpha; k) \equiv \left[\frac{2\pi}{L} \Gamma(\alpha^2) \left(\frac{L |k|}{2\pi}\right)^{1-\alpha^2}\right]^{1/2} V_j(-\alpha;k)\\
P_j^\dagger(\alpha; k) \equiv \left[\frac{2\pi}{L} \Gamma(\alpha^2) \left(\frac{L |k|}{2\pi}\right)^{1-\alpha^2}\right]^{1/2} V_j(\alpha;k)
\end{eqnarray}
are conveniently normalized Fourier transformed vertex operators. 
$\omega_k(B = \infty)$ is given by \cite{Kehrein99,Kehrein01}:
\begin{equation}
\omega_k(B=\infty)=-g_0^2\,\frac{\cos(\pi\alpha^2)}{2\Gamma^2(\alpha^2)}\,k\,|ak|^{2(\alpha^2-2)}
\label{omegak}
\end{equation}    

\section{Implementation of the forward-backward scheme}

The flow equation approach to the quantum sine-Gordon model can be used to study the real time dynamics after a sudden
interaction quench. The general idea proposed in \cite{Hackl08} consists of three steps. First, the observable
is transformed into the diagonal basis of the Hamiltonian, that is one carries
out the same sequence of infinitesimal unitary transformations:
\begin{equation}
\frac{d {\mathcal O}(B)}{d B} = [\eta(B), {\mathcal O}(B)]
\end{equation}
This is the so-called {\it forward transformation} of the observable, and in most cases it implies a very
complicated structure of $O(B=\infty)$. This observation is familiar from exact Bethe ansatz solutions.

The advantage of working in the diagonal basis comes from the actual time evolution, 
which is much easier for a diagonal Hamiltonian. Later we will see that in our model the term
${\mathcal H}_{diag}$ from Eq.~(\ref{defHdiag}) can be neglected compared to ${\mathcal H}_0$ for not too long times.
Hence, time evolution translates into phase factors and ${\mathcal O}(B=\infty, t)$ is obtained easily.

The final step is the {\it backward transformation}, where the flow of the transformed and time-evolved operator back to the original basis is carried out. The result ${\mathcal O}(t)$ is then an approximate solution to the Heisenberg equations of motion for the operator and it is straightforward to work out its expectation value with respect to
the initial (non-equilibrium) state. 

This sequence of transformations constitutes the flow equation {\it forward-backward} scheme.  It has been already used to study the real time dynamics of a Fermi liquid after a sudden interaction quench \cite{Moeckel08,Moeckel09} and the real time dynamics after a sudden quench in the ferromagnetic Kondo Model \cite{Hackl09,Kehrein09}. One of the main advantages of this approach is that it avoids the infamous problem of secular terms in perturbation theory: in time-dependent perturbative expansions, these can restrict the perturbative solution to time scales shorter than [{\it coupling constant}]$^{-1}$. In this sense, the forward-backward scheme is the quantum version of unitary perturbation theory in classical mechanics.
 
In this paper we are mainly interested in the time evolution of the bosonic number operator after a sudden
interaction quench. Hence, our first goal is the implementation of the {\it forward-backward} scheme for the creation/anhinilation operators. 

\subsection{{\it Forward} transformation}

The forward transformation requires the solution of the flow equation
\begin{equation}
\frac{d a_{i,k}(B)}{d B} = [\eta(B), a_{i,k}(B)]
\label{flow_a}
\end{equation}
with the initial condition $a_{i,k}(B=0)=a_{i,k}$.
Here $i = l,r$ corresponds to left and right movers. In order to get closed equations we make an ansatz:
\begin{eqnarray}
a_{l,k}(\infty) = h_{l,k}^{(l)}(B) a_{l,k} + h_{r,k}^{(l)}(B) a_{r,-k}^\dagger  \\ +4 \pi^2 \sum_p \Omega_k^{(l)}(p;B)\left(V_l(\alpha, p-k) V_r(-\alpha; p) - V_r(\alpha;-p) V_l(-\alpha;k-p)\right) \nonumber 
\end{eqnarray}
and likewise for the right movers. The ansatz is parametrized by various functions that must be calculated by working out the commutators in equation (\ref{flow_a}). In the weak-coupling phase this ansatz becomes exact 
in the infrared (low-energy) limit since the coupling constant $g(B)$ flows to zero.

It is convenient to decompose the transformation in two stages given by $\eta^{(1)}(B)$ and $\eta^{(2)}(B)$. 
The lowest order contribution from the first part of the generator yields the following flow equation:
\begin{equation}
\frac{d \Omega_{k}^{(l)}(p;B)}{dB} = \frac{2\alpha}{\sqrt{k}} \left(h_{l,k}^{(l)}(B) + h_{r,k}^{(l)}(B)\right) p\, u(p;B) \label{Omega_flow}   
\end{equation} 
Now we focus on the weak-coupling limit. Since the coupling constant $g(B)$ flows to zero, we can use it as a perturbative parameter in the flow equations. It can be shown that in order to preserve the bosonic commutation relations during the flow
\begin{equation}
[a_{i, k}(B), a^{\dagger}_{j, k'}(B)] = \delta_{i,j} \delta_{k,k'} \frac{L}{2\pi}
\end{equation}
it is consistent to assume an expansion of the form: $h_{l,k}^{(l)}(B) = 1 + {\mathcal O}(g^2)$ and $h_{r,k}^{(l)} = 0 + {\mathcal O}(g^2)$ (see Appendices B and C).

In the weak-coupling limit the flow of the coupling constants can be approximated by equations (\ref{weakflow1}) and (\ref{weakflow2}), which simplifies the integration of the differential equations. The result of carrying out the whole flow is:
\begin{equation}
\Omega_k^{(l)}(p;\infty) = p^{\alpha^2-3} \frac{F_\alpha}{\sqrt{k}} \, \Gamma(2-\frac{\alpha^2}{2},(2 p a)^2)  \label{Omega}
\end{equation}
where $F_\alpha = \frac{\alpha}{2\pi^2}\frac{g_0}{(2 a)^{4-\alpha^2}}(\frac{2\pi a}{L})^{\alpha^2}$. This result is valid for $\alpha^2 < 4$, which is the region we are mainly interested in. 

Now let us discuss the effect of the second part of the generator $\eta^{(2)}(B)$. Due to the structure of this generator a different approach is possible. The complete infinitesimal transformation can be rewritten as:
\begin{eqnarray}
a_{l,k}(B + dB) \simeq a_{l,k}(B) + [\eta(B), a_{l,k}(B)]dB \nonumber \\
= e^{\eta^{(2)}}\left(a_{l,k}(B) + [\eta^{(1)}(B), a_{l,k}(B)]dB \right) e^{-\eta^{(2)}}
\end{eqnarray} 
The term in brackets corresponds to the transformation already worked out above. The advantage of this expression arises from the fact that we already know the effect of the exponentiated $\eta^{(2)}(B)$ on the bosons and vertex operators:
\begin{eqnarray}
e^{\eta^{(2)}(B)} a_{l,k} e^{-\eta^{(2)}(B)} = a_{l,k}\cosh(\psi(B)) + \adr \sinh(\psi(B)) \\
e^{\eta^{(2)}(B)} a_{r,-k}^{\dagger} e^{-\eta^{(2)}(B)} = a_{r,-k}^{\dagger} \cosh(\psi(B)) + a_{l,k} \sinh(\psi(B)) \\
e^{\eta^{(2)}(B)} V_l(\alpha;p) e^{-\eta^{(2)}(B)} \simeq V_l(\alpha(1+\psi(B);p)\\
e^{\eta^{(2)}(B)} V_r(-\alpha;p) e^{-\eta^{(2)}(B)} \simeq V_r(-\alpha(1+\psi(B);p)
\end{eqnarray}
Hence the effect of the second part of the transformation is:
\begin{eqnarray}
e^{\eta^{(2)}(B)} a_{l,k}(B) e^{-\eta^{(2)}(B)} = \nonumber  \\ \left( h_{l,k}^{(l)}(B) \cosh(\psi(B)) + h_{r,k}^{l}(B) \sinh(\psi(B)) \right) a_{l,k}  \nonumber \\+ \left(h_{l,k}^{(l)} (B) \sinh(\psi(B)) + h_{r,k}^{(l)}(B) \cosh(\psi(B)) \right)\adr   \nonumber \\
+4 \pi^2 \sum_p \Omega_{k}^{(l)}(p;B) (\frac{2\pi s \sqrt{B}}{L})^{2\psi(B) \alpha^2}\nonumber \\ \times \left( V_l(\alpha\left(1+\psi(B)\right); p-k)V_r(-\alpha(1+\psi(B));p) \right. \nonumber \\- \left. V_r(\alpha \left(1 + \psi(B)\right);-p) V_l(-\alpha(1+\psi(B));k-p) \right)
\end{eqnarray}
Carrying out the whole transformation in second order of the coupling constant yields:
\begin{eqnarray}
a_{l,k}(\infty)  \simeq  \left(1 - g_0^2 \sum_p z_p^{(l)} \right) a_{l,k} + \left(g_0^2 \sum_p z_p^{(l)} + \psi(\infty)\right)a_{r,-k}^\dagger   \nonumber \\ + 4\pi^2 \sum_p \Omega_k^{(l)}(p;\infty) \left(V_l(\alpha;p-k)V_r(-\alpha;p) - V_r(\alpha;-p) V_l(-\alpha;k-p)\right) \nonumber \\
\end{eqnarray}
Here we have used a decomposition derived in Appendix B: $h_{l,k}^{(l)}(\infty) \simeq 1 - h_{r,k}^{(l)}(\infty) \simeq 1 - g_0^2 \sum_p z_p^{(l)}$. We will later see that in the present order of the calculation 
the contribution coming from the second part of the transformation can be neglected
since $\psi(\infty)\propto g_0^2$. This is consistent with our assumption that we neglect the flow of
$\alpha^2(B)$ in the weak-coupling limit: neglecting the flow of $\alpha^2(B)$ in fact just corresponds to neglecting
the generator part $\eta^{(2)}$.

\subsection{Time evolution in the diagonal basis}

The second step in the forward-backward scheme is the time evolution of the observable in the diagonal basis. Here, however, an additional approximation is required in order to solve the time evolution problem: in the diagonal Hamiltonian (\ref{diagonal}) only the bosonic kinetic term ${\mathcal H_0}$ is taken into account. We will later see that this approximation implies a maximum time scale up to which our calculation can be trusted. The time evolution dictated by ${\mathcal H}_0$ is straightforward due to the simple transformation of the vertex operators:
\begin{eqnarray}
e^{i {\mathcal H}_0 t} V_l (-\alpha, p) e^{-i {\mathcal H}_0 t} = e^{-i p t} V_l(-\alpha, p)\\
e^{i {\mathcal H}_0 t} V_r (-\alpha, p) e^{-i {\mathcal H}_0 t} = e^{i p t} V_r(-\alpha,p)
\end{eqnarray}
Therefore the time evolved anhinilation operator in the diagonal basis reads:
\begin{eqnarray}
a_{l,k}(\infty,t) = \left(1 - g_0^2 \sum_p z_p^{(l)}\right) e^{-i k t} a_{l,k} + \left(g_0^2 \sum_p z_p^{(l)} + \psi(\infty)\right) e^{i k t} a_{r,-k}^\dagger  \nonumber \\ + 4 \pi^2 e^{- i k t}  \sum_p \Omega_k^{(l)}(p;\infty) e^{2 i p t}\big(V_l(\alpha, p-k) V_r(-\alpha; p)  \nonumber \\ \qquad - V_r(\alpha; -p) V_l(-\alpha;k-p) \big) \label{time-evolved} 
\end{eqnarray}

\subsection{{\it Backward} transformation}

The final step of the transformation requires to undo the flow equation transformation for the time-evolved operator (\ref{time-evolved}). 
This is straightforward due to the perturbative nature of the transformation.
We make a general ansatz for the operator:
\begin{eqnarray}
a_{l,k}(B, t) = h_{l,k}^{(l)}(t;B) e^{- i k t} a_{l,k} + h_{r,k}^{(l)}(t;B) e^{i k t} a_{r,-k}^\dagger \nonumber \\ + 4 \pi^2 e^{-ikt}  \sum_p \Omega_{k}^{(l)}(p,t;B) e^{2ipt}\big(V_l(\alpha; p-k) V_r(-\alpha; p) \nonumber \\ \qquad - V_r(\alpha;-p) V_l(-\alpha;k-p)\big)
\end{eqnarray}
with initial conditions $h_{l,k}^{(l)}(t;\infty) = 1 - g_0^2 \sum_p z_p^{(l)}$, $h_{r,k}^{(l)}(t;\infty) = g_0^2 \sum_p z_p^{(l)}+\psi(\infty)$ and $\Omega_{k}^{(l)}(p,t;\infty) = \Omega_k^{(l)}(p;\infty)$. The flow equations resemble those for the forward transformation. since the contributions from the other functions in the ansatz follow directly via the bosonic commutation relation, we will only explicitly write down the flow equation for the function $\Omega_k^{(l)}(p,t;B)$ : 
\begin{equation}
\frac{d \Omega_{k}^{(l)}(p,t;B)}{d B} = \frac{2 \alpha}{\sqrt{k}} p u(p;B) e^{-2 i p t}
\end{equation}  
with the solution:
\begin{equation}
\Omega_k^{(l)}(p;t;0) = \Omega_k^{(l)}(p;\infty)(1-e^{-2ipt})
\end{equation}
By applying the first part of the generator the final time-evolved operator in second order of the renormalized coupling constant then reads:
\begin{eqnarray}
a_{l,k}(t) = \left(1 -  g_0^2 \sum_p (1 - e^{2 i p t}) z_p^{(l)} \right)e^{- i k t} a_{l,k} \nonumber \\ + \left(g_0^2   \sum_p (1 - e^{2 i p t} )z_p^{(l)} + \psi(\infty) \right) e^{i k t} a_{r,-k}^\dagger \nonumber \\
+ 4 \pi^2  e^{-i k t} \sum_p \Omega_k^{(l)}(p;\infty)(e^{2ipt}-1) \big(V_l(\alpha,  p-k) V_r(-\alpha; p)   \nonumber \\ \qquad -  V_r(\alpha; -p) V_l(-\alpha; k-p) \big)
\label{Heisenberga}
\end{eqnarray} 
This is the main technical result of our paper, which can be used as a building block to study the time evolution of all other observables.

\subsection{Consistency check: Ground state energy in perturbation theory}

As a consistency check for the flow equation calculation we now compare the flow equation result for the ground
state energy $E_0^{(2)}$ with second order perturbation theory.
In fact it is sufficient to evalute the kinetic energy $E_{K,0}^{(2)}$ in the ground state since one can easily 
prove $E_{K,0}^{(2)} = - E_0^{(2)}$. Our goal is therefore to calculate
\begin{equation}
E_{K,0} = \langle \bar{0}|{\mathcal H}_K|\bar{0}\rangle = \sum_k k \left(\langle \bar{0}|n_{l,k}|\bar{0}\rangle + \langle \bar{0}|n_{r,k}|\bar{0}\rangle\right)
\end{equation}
where $|\bar{0}\rangle$ is the ground state of the {\it interacting} model. 

Within the flow equation formalism this is most conveniently evaluated in the diagonal basis with the forward transformed operators
\begin{equation}
E_{K,0}^{feq} = \sum_k k \left( \langle 0|a_{l,k}^\dagger(\infty) a_{l,k}(\infty)|0\rangle 
+ \langle 0|a_{r,k}^\dagger(\infty) a_{r,k}(\infty)|0\rangle \right)
\end{equation}  
where we have used $\langle \bar{0}|n_{i,k}|\bar{0}\rangle = \langle 0|n_{i,k}(B=\infty)|0\rangle$. 
Here $|0\rangle$ is the bosonic vacuum since this is trivially the ground state in the diagonal basis.
If this calculation is carried out using our previous results from the flow equation formalism, the result actually does not coincide with perturbation theory. However, this is simply due to the fact that the flow equation calculation is a renormalized expansion,
whereas conventional perturbation does not contain renormalization effects. 
In order to compare with perturbation theory\footnote{And only for that reason since renormalization effects due to
the running coupling constant are essential in the sine-Gordon model.} we therefore artifically set $g(B) = g_0$
and find:
\begin{equation}
E_{K,0}^{feq} = \frac{g_0^2}{a} \frac{L}{2\pi a}\frac{\Gamma(2\alpha^2-2)}{\Gamma^2(\alpha^2)}
\label{EK0feq}
\end{equation}
The same result can be obtained by working out the kinetic energy in second order perturbation theory:
\begin{equation}
E_{K,0}^{(2)} = \int_0^\infty d\lambda \sum_{n\neq0}|\langle n|{\mathcal H}_I|0\rangle|^2e^{-\lambda \sum_k k(n_{l,k} + n_{r,k})}
\end{equation}
where $|n\rangle \equiv |n_{l,k_1},n_{l,k_2},...,n_{r,k_1}, n_{r,k_2},..\rangle$. By using the matrix elements of vextex operators given in Appendix A we arrive at the following expression:
\begin{equation}
E_{K,0}^{(2)} = 2 L \frac{g_0^2}{(2\pi a^2)^2} \left(\frac{2\pi a}{L}\right)^{2\alpha^2} \int dx \int_0^\infty d\lambda e^{2\alpha^2\sum_k \frac{1}{k} \cos(kx) e^{-\lambda k}}
\end{equation}
After working out the integrals it can be shown that the flow equation result (\ref{EK0feq}) is reproduced.

\section{Real time dynamics after a sudden interaction quench}

We now want to use the previous results to investigate the non-equilibrium dynamics of the
quantum sine-Gordon model in the weak-coupling limit. Specifically, we study the mode 
occupation numbers after a sudden interaction quench:
\begin{equation}
{\mathcal H}(t) = {\mathcal H}_0 + \Theta(t) {\mathcal H}_I
\end{equation}
Since the system is prepared in the non-interacting ground state $|0\rangle$ (bosonic vacuum)
of ${\mathcal H}_0$ for $t<0$, the time-dependent occuptation number for left movers is simply 
\begin{eqnarray}
\langle n_{l,k} \rangle(t) = \langle 0 (t) | n_{l,k}| 0 (t) \rangle = \langle 0| a_{l,k}^\dagger(t) a_{l,k}(t) | 0 \rangle \label{time_n}
\end{eqnarray}
where we insert the time-evolved operators in the Heiseberg picture (\ref{Heisenberga}).
It is this conceptual simplicity which makes sudden interaction quenches
very appealing for studying non-equilibrium problems.
Notice that an identical result to (\ref{time_n}) can be obtained for right movers, and therefore
we restrict ourselves to explicit expressions for left movers only. 

\begin{figure}
\begin{center}
\includegraphics[width=4in]{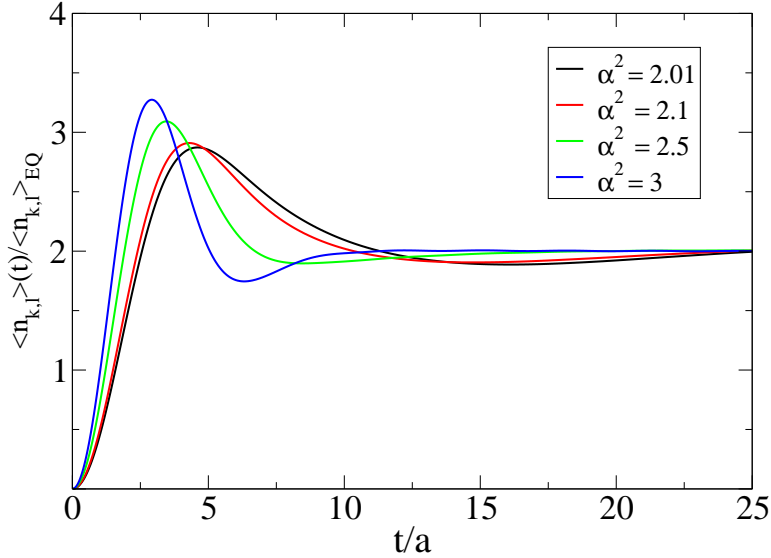}
\caption{Real time dynamics of the occupation number normalized to the equilibrium value for different values of $\alpha^2$
after an interaction quench. All the curves correspond to the same momentum $k\, a = 0.1$. After a few oscillations the occupation numbers converge to twice their equilibrium values, see text.}
\label{fig3}
\end{center}
\end{figure}

An expression for (\ref{time_n}) can be worked out readily from the result (\ref{Heisenberga}) obtained within the forward-backward scheme:
\begin{equation}
\langle n_{l,k}(t) \rangle = \frac{\alpha^2 g_0^2}{k}\frac{L}{2\pi a}\frac{4^{\alpha^2-2}}{(\Gamma(\alpha^2))^2} I(ka, \frac{t}{a}) +O(g_0^4)
\end{equation}
with the integral
\begin{eqnarray}
I(ka,\frac{t}{a}) = \int_{ka}^{\infty} dx \sin^2(x \frac{t}{a})(\Gamma(2-\frac{\alpha^2}{2},4x^2))^2 x^{3\alpha^2-7}(ka - x)^{\alpha^2-1} e^{-x} \nonumber
\end{eqnarray}
In order to get this result, we have made use of the properties of vertex operators summarized in Appendix A and of the expression for $\Omega_k^{(l)}(p;\infty)$ given in (\ref{Omega}). However, it is more convenient to express this result in terms of the equilibrium occupation numbers. Fortunately, the equilibrium occupation numbers follow directly from the flow equation calculation:
\begin{eqnarray}
\langle n_{l,k} \rangle_{EQ} = \langle \bar{0}| n_{l,k} |\bar{0} \rangle = \langle 0 | a_{l,k}^\dagger(B=\infty) a_{l,k}^\dagger (B=\infty) |0\rangle  \nonumber \\ = \frac{\alpha^2 g_0^2}{k}\frac{L}{2\pi a}\frac{4^{\alpha^2-2}}{(\Gamma(\alpha^2))^2} I_{EQ}(ka) +O(g_0^4)
\end{eqnarray}
Here $|\bar{0}\rangle$ denotes the interacting ground state, and:
\begin{equation}
I_{EQ}(ka) = \frac{1}{4} \int_{ka}^{\infty} dx (\Gamma(2-\frac{\alpha^2}{2},4x^2))^2 x^{3\alpha^2-7}(k a - x)^{\alpha^2-1} e^{-x}
\end{equation}

\begin{figure}
\begin{center}
\includegraphics[width=4in]{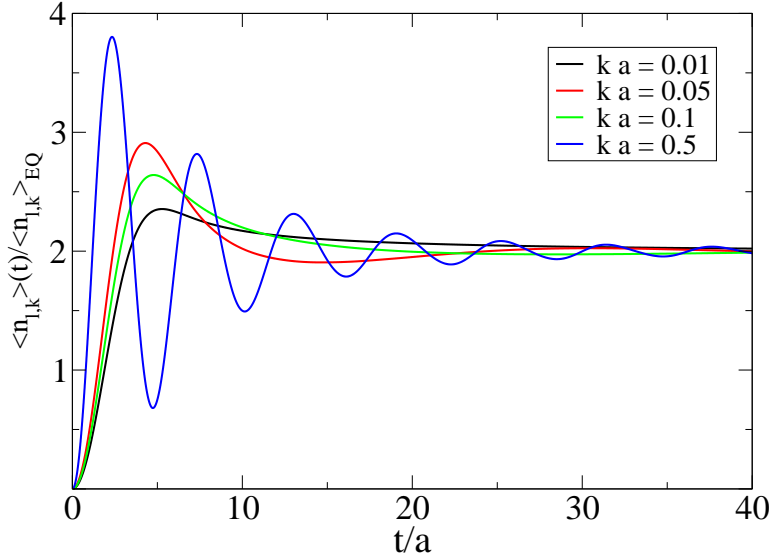}
\caption{Real time dynamics of the mode occupation number normalized to the equilibrium value for different momenta $k\, a$ and
fixed $\alpha^2 = 2.1$. The oscillations disappear in the infrared limit.}
\label{fig4}
\end{center}
\end{figure}

This results in the following compact expression for the ratio of non-equilibrium to equilibrium occupation number:
\begin{equation}
\frac{\langle n_{l,k}(t) \rangle}{\langle n_{l,k} \rangle_{EQ}} =   \frac{I(ka,\frac{t}{a})}{I_{EQ}(ka)} \: +O(g_0^2)
\end{equation}
and likewise for right movers. 
The integrals in this expression must be computed numerically. Plots of these ratios are shown in Fig. \ref{fig3} for different values of $\alpha^2$ and fixed momentum, and in Fig. \ref{fig4} for fixed $\alpha^2$ and different momenta $k\, a$. 

One observes that the key phenomena after the quench are
damped oscillations of the mode occupation on a time scale set by the lattice cutoff $a$. 
The asymptotic value of the mode occupation universally converges to twice its equilibrium value: 
\begin{equation}
\langle n_{l,k}(t\rightarrow\infty) \rangle = \left(2+O(g_0^2)\right) \langle n_{l,k} \rangle_{EQ} 
\label{factor2}
\end{equation}
This can be understood easily by noticing that $I_{EQ}(ka)$ and $I(ka,t/a)$ only
differ by replacing a factor $1/4$ by $\sin^2(xt/a)$ in the integrand. Clearly
the limit $\langle n_{l,k}(t\rightarrow\infty) \rangle$ just amounts to taking the
time average over $\sin^2(xt/a)$ in the integrand, which gives $1/2$ and therefore
$I(ka,t/a\rightarrow\infty)=2\,I_{EQ}(ka)$.

Notice that (\ref{factor2}) implies a non-thermal mode distribution function for the asymptotic state
of this closed quantum system: the equilibrium system with nonzero
temperature cannot reproduce this expression.

\section{Conclusions}

We have demonstrated that an interaction quench in the weak-coupling phase of the
sine-Gordon model leads to an interesting dynamics that is quite different from a
quench of the forward scattering only. Quenching the forward scattering does not
induce any dynamics for the bosonic occupation numbers, which are in fact constants
of motion in this case \cite{Cazalilla06}. On the other hand, we have found that
a quench of the backscattering term in the weak-coupling phase leads to a real time
dynamics where the excitation energy of the quench is converted into bosonic mode
occupations that oscillate on a time scale set by the ultraviolet cutoff and eventually
reach twice their equilibrium values (\ref{factor2}). This factor~2 is {\it universal}
for weak interaction quenches and has previously also been seen for the momentum
distribution function in the non-equilibrium
Hubbard model \cite{Moeckel08,Moeckel09} and for the magnetization of the non-equilibrium
ferromagnetic Kondo model \cite{Hackl09,Kehrein09}. It occurs for quenches in quantum
systems where i)~second order perturbation theory is valid (at least up to a certain time scale)
and ii)~for observables like the mode occupation number operator which commute with 
${\mathcal H}_0$ (for a proof and more details on the conditions see Ref.~\cite{Moeckel09}). 

Since such general results are important for developing a better understanding of
non-equilibrium quantum many-body systems in general, we need to critically re-examine
the approximations in our calculation. Due to the weak-coupling behavior of the running
coupling constant, the second order calculation presented here becomes more and more
reliable in the infrared limit. Therefore higher order corrections to the universal factor~2
will vanish in the low-energy limit. However, we had to make the additional approximation
to neglect the time evolution generated by ${\mathcal H}_{diag}(\infty)$ from (\ref{defHdiag}).
Therefore the result (\ref{factor2}) can only be trusted up to the time scale
\begin{equation}
\tau_k\propto \omega_k^{-1}(B=\infty) \propto g_0^{-2}\,k^{3-2\alpha^2}
\end{equation}
Whether our central result
\begin{equation}
\langle n_{i,k}(t\rightarrow\infty) \rangle = 2 \langle n_{i,k} \rangle_{EQ} 
\label{factor2concl}
\end{equation}
really holds beyond the time scale $\tau_k$ or only in a time window $a\ll t\ll\tau_k$
cannot be answered based on our calculation. The interaction quench in the weak-coupling
phase of the sine-Gordon model therefore occupies an interesting place between the
ferromagnetic Kondo model, where the factor~2 is in fact asymptotically exact \cite{Hackl09,Kehrein09},
and the non-equilibrium Hubbard model in $d\geq 2$ dimensions, where the factor~2
describes the prethermalization regime \cite{Moeckel08,Moeckel09} before the system
eventually thermalizes. The integrability of the sine-Gordon model could make one
expect that the non-equilibrium distribution function (\ref{factor2concl}) remains stable
for all times and does not approach a thermal limit form. However, the conserved
quantities in the sine-Gordon model do not impose any obvious constraints on the dynamics
of the momentum distribution function, unlike in the case of quenching the forward 
scattering \cite{Cazalilla06}. 

Further studies of this question would be very worthwile, either numerical or based on
an exact solution. We have seen that the weak-coupling quench in the sine-Gordon model
is on the borderline between thermalization and non-thermalization seen through
the eyes of the mode distribution function. This is similar to the role played 
by the celebrated Fermi-Pasta-Ulam problem for classical many-body systems \cite{FPU,FPUReview}.
A better understanding of the weak-coupling quench in the quantum sine-Gordon model
could be an important step in elucidating the fundamental question of thermalization
in the quantum world.

\vspace*{1cm}
\noindent
{\bf Acknowledgments}
\newline
This work was supported by the Deutsche Forschungsgemeinschaft (DFG) through 
FG 960. S.~K. also acknowledges support through
the Center for Nanoscience (CeNS) Munich and the German Excellence Initiative via the Nanosystems Initiative
Munich (NIM). J.~S. was supported by MEC (Spain) through grants FIS2007-65723, FIS2008-00124 and CONSOLIDER CSD2007-00010, and by the Comunidad de Madrid through CITECNOMIK. J.~S. also wants to acknowledge the I3P Program from the CSIC for funding.

\appendix

\section{Properties of vertex operators}

We summarize some important properties of vertex operators ($k,k'>0$). More details can be found in Ref.~\cite{Kehrein01}.
\begin{eqnarray}
V_l(\alpha;-k)|0\rangle = V_l(-\alpha;k)|0\rangle = V_r(\alpha;k)|0\rangle = V_r(-\alpha;-k)|0\rangle = 0 \nonumber \\\\
\langle 0| V_l(-\alpha; k) V_l(\alpha; k')|0\rangle = \langle 0|V_r (\alpha; k) V_r(-\alpha; k') |0\rangle  \nonumber \\ =\delta_{k,k'} \Theta(k) \left(\frac{L}{2\pi}\right)^{\alpha^2+1}\frac{|k|^{\alpha^2-1}}{\Gamma(\alpha^2)} \\
\langle 0| V_l(\alpha; k) V_l(-\alpha; k')|0\rangle = \langle 0|V_r (-\alpha; k) V_r(\alpha; k') |0\rangle  \nonumber \\ =\delta_{k,k'} \Theta(-k) \left(\frac{L}{2\pi}\right)^{\alpha^2+1}\frac{|k|^{\alpha^2-1}}{\Gamma(\alpha^2)} \\
\langle 0| V_l(\alpha; k) V_l(\alpha; k')|0\rangle = \langle 0|V_r (\alpha; k) V_r(\alpha; k') |0\rangle = 0 
\end{eqnarray}
The operator product expansion of left handed vertex operators reads:
\begin{eqnarray}
 \, 
*V_l(-\alpha;k) V_l(\alpha; k')* = \frac{\alpha}{\Gamma(\alpha^2-1)}\left(\frac{L}{2\pi}\right)^{\alpha^2}\nonumber \\ \times\left(\sqrt{k'-k}|k|^{\alpha^2-2} \theta(k) \theta(k'-k) a_{l,k'-k}^\dagger \right. \nonumber \\ \left. + \sqrt{k-k'}|k'|^{\alpha^2-2} \theta(k') \theta(k-k') a_{l,k-k'} + ...\right) \label{OPE1}\\
\, *V_l(\alpha;k) V_l(-\alpha;k')* = - *V_l(-\alpha;-k) V_l(\alpha;-k')* \label{OPE2}
\end{eqnarray}
Matrix elements of vertex operators between number states are:
\begin{eqnarray}
\langle n | V_l(\alpha;x)|0\rangle = \Pi_{k>0} \left(\frac{\alpha}{\sqrt{k}}\sqrt{\frac{2\pi}{L}}\right)^{n_{l,k}} \frac{e^{-i k n_{l,k} x}}{\sqrt{n_{l,k}}} \\
\langle n | V_r(\alpha;x)|0\rangle = \Pi_{k>0} \left(\frac{\alpha}{\sqrt{k}}\sqrt{\frac{2\pi}{L}}\right)^{n_{r,k}} \frac{e^{i k n_{r,k} x}}{\sqrt{n_{r,k}}} 
\end{eqnarray}

\section{Flow equation for $h_{l,k}^{(l)}(B)$}
The flow equations for the coefficient $h_{l,k}^{(l)}(B)$ can be obtained from the commutator of the generator with the second part of the ansatz. Only the terms linear in $a_{l,k}$ are kept, and they come from the operator product expansion of two vertex operators:
\begin{eqnarray}
-4(2\pi)^4 \sum_{p,p'} p u(p;B) \Omega_k^{(l)}(p';B) \nonumber \\ \times \left( *V_l(\alpha;p) V_l(-\alpha;k-p')* \langle 0| V_r(-\alpha;p)V_r(\alpha;-p')|0\rangle \right. \nonumber\\ \left. + *V_l(-\alpha;p) V_l(\alpha; p'-k)* \langle 0| V_r(\alpha;-p') V_r(-\alpha;p) |0\rangle \right) 
\end{eqnarray}
By using the expressions given in Appendix A one finds the flow equation:
\begin{eqnarray}
\frac{ d h_{l,k}^{(l)}(B)}{d B} = - \alpha \sqrt{k} \frac{4 (2 \pi)^4}{\Gamma(\alpha^2) \Gamma(\alpha^2-1)}\left(\frac{L}{2\pi}\right)^{2\alpha^2} \nonumber \\ \times \left(\sum_{p>0} u(p;B) \Omega_k^{(l)}(p;B) p^{2\alpha^2-2} + \sum_{p>k} u(p;B) \Omega_k^{(l)}(p;B) p^{\alpha^2}|p-k|^{\alpha^2-2} \right) \nonumber \\
\end{eqnarray}
which is in second order in the coupling constant $g_0$. The integration of this equation turns out to be easy by using the flow equation (\ref{Omega_flow}) for the parameter $\Omega^{(l)}_{k,l}(B)$. At leading order in the coupling constant the solution reads:
\begin{eqnarray}
h_{l,k}^{(l)}(B) = 1 - \frac{(2\pi)^4 k}{\Gamma(\alpha^2) \Gamma(\alpha^2-1)}\left(\frac{L}{2\pi}\right)^{2\alpha^2} \nonumber \\ \times  \left(\sum_{p>0} (\Omega_k^{(l)})^2(p;B) p^{2\alpha^2-2} +  \sum_{p>k} (\Omega^{(l)}_k)^2(p;B) p^{\alpha^2} (p-k)^{\alpha^2-2}\right)
\end{eqnarray}
We get (for $k a \ll 1$):
\begin{eqnarray}
h_{l,k}^{(l)}(B) = 1 - \frac{2 (2\pi)^4}{\Gamma(\alpha^2) \Gamma(\alpha^2-1)}\left(\frac{L}{2\pi}\right)^{2\alpha^2}\sum_{p>0} \Omega^2(p;B) p^{2\alpha^2-2}
\end{eqnarray}
where we have defined $\Omega_k^{(l)}(p;\infty) \equiv \Omega^{(l)}(p;\infty) / \sqrt{k}$.
A similiar derivation can be done for the right-handed operators yielding $h_{r,k}^{(l)}(B) \simeq 1 - h_{l,k}^{(l)}$.

\section{Sum rule}

The canonical commutation relations must be fulfilled during the entire flow:
\begin{equation}
[a_{l,k}(B), a_{l,k'}^\dagger(B)] = \frac{L}{2\pi} \delta_{k,k'}
\end{equation}
From this requirement one can derive a consistency condition for the flowing operator. Evaluating this condition in the ground state gives:
\begin{eqnarray}
\delta_{k,k'} = h_{l,k}^2 \delta_{k,k'} - h_{r,k}^2 \delta_{k,k'} + \frac{(2\pi)^4}{\Gamma(\alpha^2)^2} \left(\frac{L}{2\pi}\right)^{2\alpha^2}\nonumber \\ \times \left(\sum_{p>0} \Omega_k^2(p;B) p^{\alpha^2} (p+k)^{\alpha^2-1} - \sum_{p>k} \Omega_k^2(p;B) p^{\alpha^2} (p-k)^{\alpha^2-1}\right) 
\end{eqnarray}
Using some straightforward algebra one can verify that this is indeed fulfilled for the flow equations 
derived in this paper.

\end{document}